\documentstyle[preprint,aps,epsf,graphicx,floats]{revtex}    
\def\bq{\begin{equation}}
\def\eq{\end{equation}}
\def\ba{\begin{eqnarray}}
\def\ea{\end{eqnarray}}

\begin{document}
\thispagestyle{empty}

\newcommand{\sla}[1]{/\!\!\!#1}

\preprint{
\font\fortssbx=cmssbx10 scaled \magstep2
\hbox to \hsize{
\hskip.5in \raise.1in\hbox{\fortssbx Fermilab}
\hfill\vtop{\hbox{\bf }
            \hbox{FERMILAB-Pub-00/146-T}
            \hbox{December 1, 2000}} }
}

\title{
$p\bar{p}\to t\bar{t}H$: \\
A Discovery mode for the Higgs boson at the Tevatron
}
\author{J.~Goldstein$^1$, C.~S.~Hill$^2$, J.~Incandela$^1$, Stephen Parke$^3$, 
D.~Rainwater$^3$ and D.~Stuart$^1$\\[3mm]}
\address{
$^1$Particle Physics Division, 
Fermi National Accelerator Laboratory, 
Batavia, IL 60510, U.S.A.\\
$^2$Department of Physics, University of California, Davis, CA 95616, U.S.A.\\
$^3$Theoretical Physics Department, Fermi National Accelerator Laboratory,\\ 
Batavia, IL 60510, U.S.A.
}
\maketitle
\begin{abstract}
The production of a Standard Model Higgs boson in association with a top 
quark pair at the upcoming high luminosity run (15 fb$^{-1}$ integrated 
luminosity) of the Fermilab Tevatron ($\sqrt{s} = 2.0$ TeV) is revisited.
For Higgs masses below 140 GeV we demonstrate that the production cross 
section times branching ratio for $H\to b\bar{b}$ decays yields a 
significant number of events and that this mode is competitive with and 
complementary to the searches using $p\bar{p} \to WH,ZH$ associated 
production. For higher mass Higgs bosons the $H\to W^+W^-$ decays are 
more difficult but have the potential to provide a few spectacular events.
\end{abstract}

\vspace{3mm}

One of the untested features of the elegant and highly successful Standard Model 
(SM) of electroweak interactions is the Higgs mechanism for spontaneous 
electroweak symmetry breaking. 
The Higgs sector of the SM consists of one doublet of complex scalar fields 
which is used to break the SU(2)$\times$U(1) electro-weak 
gauge symmetry to U(1) electromagnetism. 
This mechanism gives rise to the masses of the W and Z bosons and is also used
to give masses to the charged leptons and quarks.
An additional physical neutral scalar particle is also produced, the
undiscovered Higgs boson.
All properties of this boson are predicted by the SM except for its mass.
Most extensions to the SM's electroweak symmetry breaking mechanism
predict more than one physical scalar state, 
however one of these states tends to have properties very similar to
the SM Higgs boson.

Direct collider searches have so far ruled out a SM Higgs boson with a mass 
less than $\approx 107$~GeV~\cite{LEP_limits} at the $95\%$ confidence level. 
The final run of the Large Electron Positron collider (LEP) at CERN this year 
should be able to extend the exclusion limit 
or provide a $5\sigma$ discovery up to about 114~GeV. 
The next CERN collider, after LEP, 
is the 14 TeV proton-proton Large Hadron Collider (LHC), which will have the 
capability to discover a SM Higgs boson of nearly any 
mass\footnote{In the SM $M_H \gtrsim 1$~TeV is ruled out 
by unitarity constraints.} in multiple search channels.
However, the LHC is not scheduled to begin taking data until 2005 at the 
earliest. 
In the meantime, the upcoming Run II of Fermilab's 2 TeV proton-antiproton 
collider, the Tevatron, will have an opportunity to discover a Higgs boson 
above the LEP exclusion limit and below 180 GeV. 
In this Letter we will concentrate on the SM case over this mass range.

The SM Higgs boson couples at lowest order to all SM particles with a 
strength proportional to their mass.
For the massless photon and gluon a coupling is induced at higher order 
via loops of heavy particles.
Therefore there are many ways for the Higgs boson to decay and also
many ways to produce it at a hadron collider.
For the mass range of interest here, 
$100 \lesssim M_H \lesssim 200$~GeV, the Higgs boson decays 
predominantly to either a pair of $b$-quarks ($b\bar{b}$) for a mass
less than 135 GeV, or to four fermions via real or virtual $W$ gauge bosons 
for a mass above 135 GeV. 
These decays are referred to as $H\to b\bar{b}$ and  
$H\to W^+W^-$ respectively, and at 135 GeV they have approximately equal
branching ratios ($43\%$ to $40\%$ respectively). 
At this mass, the remainder of the decay branching ratio is primarily into 
$\tau$ lepton, $c$ quark and gluon pairs.

\begin{figure*}[t]
\begin{picture}(0,0)(0,0)
\includegraphics{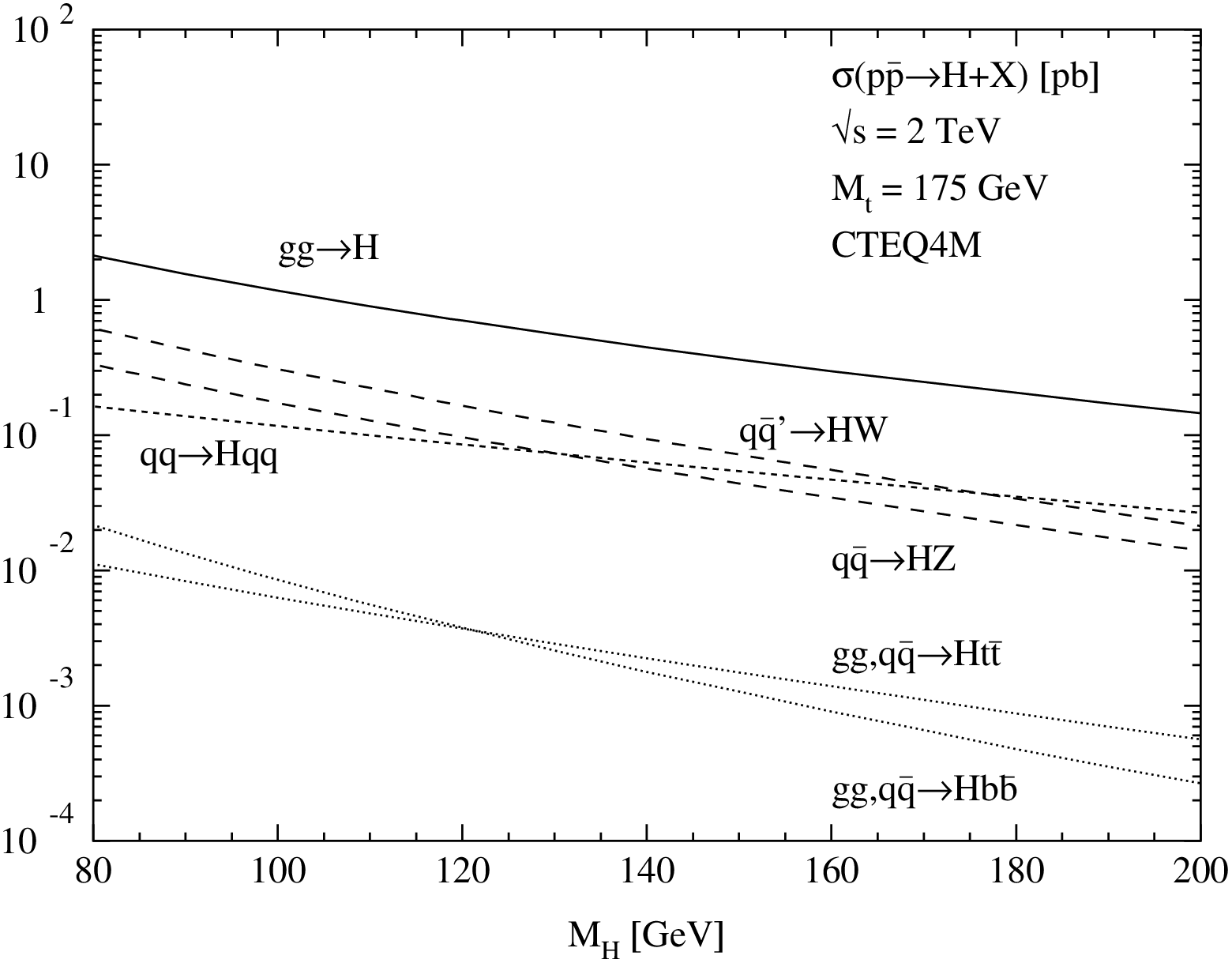}
\includegraphics{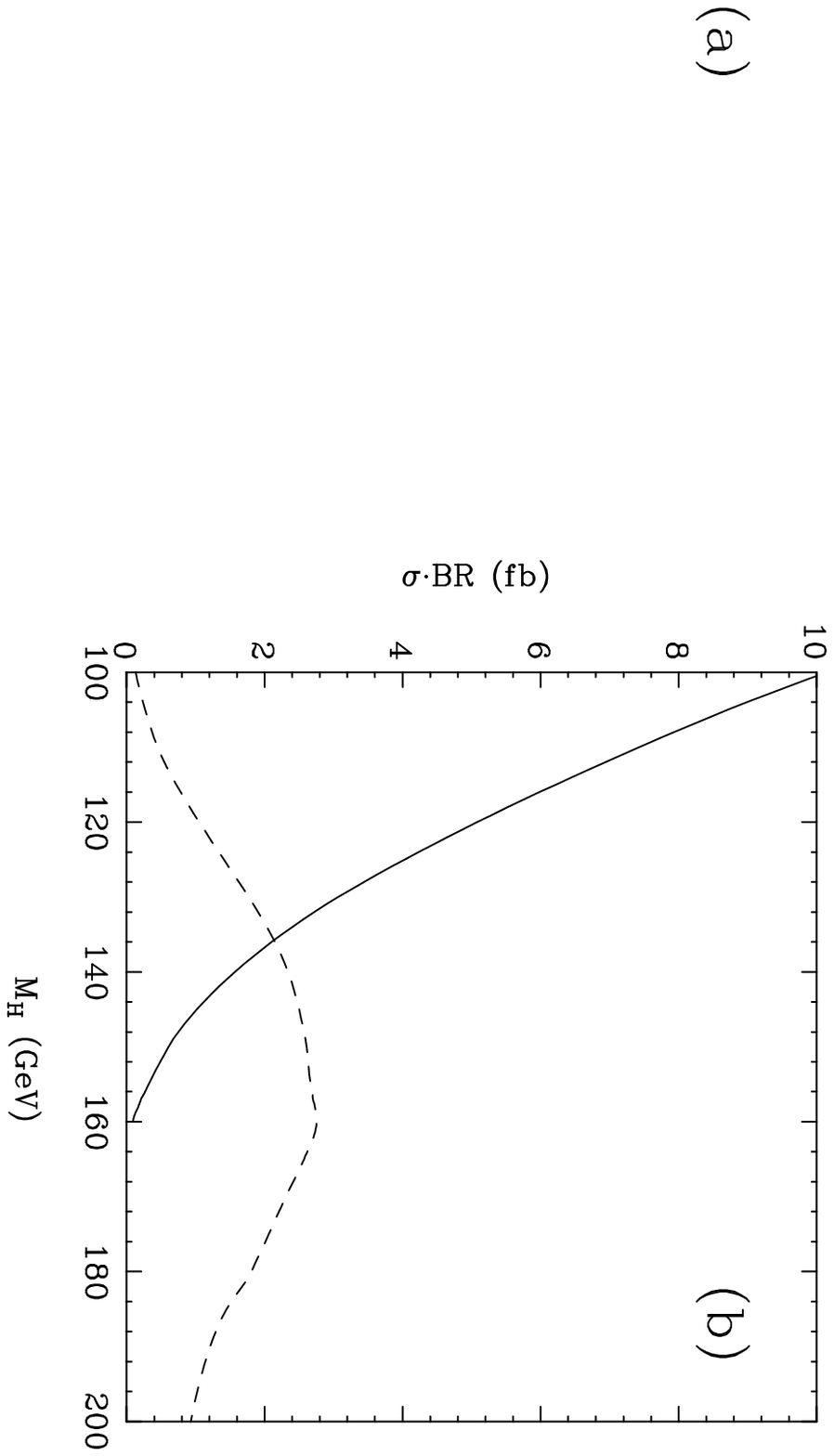}
\end{picture}
\vspace{7.5cm}
\caption{
For $p\bar{p}$ collisions at $\sqrt{s} = 2.0$~TeV: 
(a) cross section (pb) for various Higgs boson production 
modes~\protect\cite{Hobbs}; 
(b) the $t\bar{t}H$ production cross section times branching ratio 
and K-factor (fb) for $H\to b\bar{b}$ (solid) and $H\to W^+W^-$ (dashed).}
\label{fig:rates}
\end{figure*}

Cross sections for various Higgs boson production modes at the Tevatron are 
shown in Fig.~\ref{fig:rates}(a)~\cite{Hobbs}. 
The dominant mode is via gluon-gluon fusion, $gg\to H$, 
which proceeds via a top quark loop.
The next largest mode is Higgs boson associated production with a $W$ or $Z$ 
boson.
At a still lower cross-section is Higgs boson production via 
bremsstrahlung off a top quark pair.
The current searches planned for the 
Tevatron Run II concentrate on the dominant production modes: 
$WH,ZH$ associated production with decays $H\to b\bar{b}$~\cite{Marciano}, 
and gluon fusion 
production $gg\to H$ with subsequent decay $H\to W^+W^-$~\cite{Han}. 
While detailed studies with detector simulation and neural net analysis
show promise, 
they do indicate the need for large integrated luminosity, and even with 
30~fb$^{-1}$ these searches have difficulty in the mass regions 130-150~GeV 
and above 180~GeV. 
In this letter we investigate the feasibility of a search in $t\bar{t}H$ 
associated production, first proposed in Refs.~\cite{ttH} in the context of 
the Superconducting Supercollider, and later examined in more detail for the 
LHC in particular; see~\cite{ttH_detail} and~\cite{LHC_TDR}.
Decays to both $H\to b\bar{b}$ and $H\to W^+W^-$ are considered, with a 
varying number of tagged $b$ quarks and additional jets in the final state 
and at least one charged lepton\footnote{Top quarks decay predominantly to 
a $W$-boson and a $b$-quark. The $W$-boson subsequently decays
2/3 of the time to a pair of quarks (jets) and 1/9 of the time into 
each of $e\nu_e$, $\mu\nu_\mu$ and $\tau\nu_\tau$.  
The neutrinos are not observed and 
the $\tau$-lepton further decays 2/3 of the time to a jet plus neutrinos 
and 1/3 of the time to $e$ or $\mu$ and more neutrinos. Thus, about $26\%$ 
of the time a $W$-boson decays into a taggable charged lepton (e or $\mu$).} 
to help discriminate against the large QCD backgrounds.
We believe that, in spite of the low cross section, a search for the 
SM Higgs boson in association with a top quark pair in the Tevatron Run II 
data set (15 fb$^{-1}$ integrated luminosity) is likely to be comparable
to the early search and evidence for the top quark results from CDF in 
Run I~\cite{tt_run1}.

We have calculated the signal at the parton level for $p\bar{p}$ collisions at 
$\sqrt{s} = 2.0$~TeV using exact tree-level matrix elements generated by 
{\sc madgraph}~\cite{Madgraph} and {\sc comphep}~\cite{Comphep}, and NLO 
corrected decay rates of the Higgs boson via {\sc hdecay}~\cite{Hdecay}. 
We take the K-factor to be 1.33: the ratio of the NLO $t\bar{t}$ cross 
section to the leading order value. Our estimate agrees well with a calculation 
in the effective Higgs approximation~\cite{EHA}, which estimates the K-factor 
to be 1.2 to 1.5 depending on the Higgs boson mass and other uncertainties.
CTEQ4L~\cite{CTEQ} parton distribution functions are used, 
and both the factorization and renormalization scales 
are taken as the top quark mass, $m_t$. 
The cross section times Higgs branching ratio times K-factor used
throughout the rest of this letter are shown in 
Fig.~\ref{fig:rates}(b). 
For $H\to b\bar{b}$ the cross section times branching ratio falls steeply and 
becomes smaller than 1~fb for Higgs masses above 140 GeV.
In contrast, the cross section times branching ratio
for $H\to W^+W^-$ is a broad bump above 1~fb between 125 and 190 GeV.
Thus, a search strategy developed for a 160~GeV Higgs would likely work over 
a much wider mass range, covering some of the unobservable regions 
of the current searches. 

For lower Higgs boson masses ($<$ 140 GeV), the most interesting decay
mode of the Higgs is $H\to b\bar{b}$ and the 
final state of the hard interaction is $W^+W^- b\bar{b}b\bar{b}$.
For higher mass Higgs bosons, $H\to W^+W^-$ is the dominant decay mode
and the final state signature is instead $W^+W^-W^+W^- b\bar{b}$. 
For both modes, each W-boson will decay either hadronically into a quark and 
anti-quark or leptonically into a charged lepton and neutrino. 
Backgrounds to these modes consist of resonant production, 
e.g. irreducible $t\bar{t}Z$ events with $Z\to b\bar{b}$ 
or $Z\to \ell^+\ell^-$; continuum $t\bar{t}b\bar{b}$ or 
$t\bar{t}c\bar{c}$ production; and reducible 
backgrounds such as $t\bar{t} + jets$, where the additional jets are light 
quarks or gluons but may be misidentified as $b$ quarks. 
Table~\ref{bkgs} itemizes the major backgrounds and gives their 
cross sections. All other SM backgrounds are negligible. 
Clearly, the largest background arises from $t\bar{t}+jets$. 
Thus, the search for $t\bar{t}H$ will critically depend on the 
detector ability to tag $b$ quark jets and simultaneously to suppress 
mistagging of non-$b$ jets.

\begin{table*}
\caption{Inclusive cross sections for the major physics and reducible 
backgrounds for $p\bar{p}$ collisions at $\sqrt{s} = 2.0$~TeV. Continuum 
production of final states involving heavy quarks or weak bosons plus 
additional jets include a $p_T \geq 20$~GeV cut on the additional jets, but 
no cut on the decays of heavy states. Typical uncertainties on these 
estimates are on the order of $\pm 50\%$.}
\vspace{0.15in}
\label{bkgs}
\begin{tabular}{l|c||l|c}
backgrounds to $H\to b\bar{b}$ & cross section (fb) &
backgrounds to $H\to W^+W^-$   & cross section (fb) \\
\hline
$t\bar{t} + jj$	 ~~~~($\triangle R(jj) > 0.4$)	& 1030   &
$t\bar{t} + jj$					& 1030   \\
$t\bar{t} + b\bar{b}$ (or $c\bar{c}$)		&   27   &
$t\bar{t} + W$					&   17   \\
$t\bar{t} + Z, Z\to b\bar{b}$			&    1.5 &
$t\bar{t} + Z, Z\to \ell^+\ell^-$		&    0.9 \\
$WZ + jj, Z\to b\bar{b}, W\to e\nu,\mu\nu$	&   10   & & \\
\end{tabular}
\end{table*}

We have calculated the parton level cross section for $t\bar{t}+jj$ events 
using the exact tree-level matrix elements~\cite{Comphep,Stange}, 
which contain both collinear and soft singularities for the additional jets. 
The collinear singularities are removed by imposing the requirement that the 
final-state partons be well-separated in space, $\triangle R(jj) > 0.4$. 
This corresponds to the experimental requirement that the observed jets
be similarly separated.\footnote{$\triangle R$ is a separation in the space of 
detector coordinates azimuth and pseudorapidity, 
$\triangle R(jj) = \sqrt{ \triangle\phi^2 + \triangle\eta^2}$.} 
The severity of the soft singularity for a given $p_T$ cut
may be estimated by examining the ratio of the cross sections 
for $t\bar{t}+jets$ production to that for $t\bar{t}$, the latter of which may 
be regarded as an inclusive rate; the overall rate can later be normalized to 
NLO calculations. 
For the Tevatron we find $\sigma_{t\bar{t}jj}/\sigma_{t\bar{t}} \approx 1/7$, 
including a requirement of $p_T > 20$~GeV for the additional jets, 
indicating that the calculation is largely perturbative. 
This is in line with general expectations 
that each additional gluon jet of $p_T \gtrsim 20$~GeV would multiply the 
inclusive rate by a factor $\approx 1/2-1/3$. 
One may also use an 
exponentiation approximation for the soft gluons~\cite{expon}, which yields a 
result $30\%$ lower than the value in Table~\ref{bkgs}. 

We have compared the results of our matrix element calculation 
with Pythia 6.115~\cite{Pythia}, and while there are differences 
in cross section as a 
function of jet $p_T^{min}$, the differences at Tevatron energies are largely 
within the uncertainties of the matrix element calculations.
Ultimately, any uncertainty in the rate for additional jets in $t\bar{t}$ 
events is not a concern as the $t\bar{t}+jets$ sample in Run II 
will be large enough to calibrate the Monte Carlos. 
Furthermore, by retaining the larger estimate we make this analysis 
more conservative.

These uncertainties aside, it is clear that a major background could come from 
the mistagging of non-$b$ jets. Both the efficiency for $b$ jet tagging and 
non-$b$ jet suppression will thus be extremely important in this search. 
Our studies of the upgraded Tevatron collider detectors using Run I algorithms 
predict an average $b$ jet tagging efficiency of about $60\%$ and a $c$ quark 
mistag rate of about $25\%$, 
while at the same time providing at least a factor 
of 500 suppression of light quark and gluon jet mistags. 
We estimate that implementation of 3-D vertexing algorithms will improve 
the $b$ tagging efficiency to at $\sim 70\%$ while leaving the mistagging 
rates unchanged.
We use the 3-D values in our analysis ($70\%, 25\%$ and $0.2\%$).

For a Higgs boson of mass $M_H \lesssim 140$~GeV, where the 
$W^+W^- b\bar{b}b\bar{b}$ signature is preferred, 
a $70\%$ $b$ tagging efficiency translates to a $92\%$ probability to tag 
$\geq 2$ $b$ jets, $65\%$ for $\geq 3$ tags, and $24\%$ to tag all four. 
As an example, consider the case of $M_H = 120$~GeV, with one $W$ 
decaying leptonically and one decaying hadronically, which has much better 
known backgrounds than the larger all-hadronic $W$ decay channel.
The cross section times branching ratio is 2.0~fb. 
For three $b$ tags, 15~fb$^{-1}$ of data would contain $\sim 19$ signal events 
and about 140 background events. 
Requiring four $b$ tags would yield a signal sample of approximately 7 events, 
on top of a background of about 40 events. 
These counts are before any top reconstruction or Higgs boson mass binning. 

\begin{figure*}[t]
\vspace*{-0.6cm}
\begin{center}
\epsfxsize=6.5in
\epsfysize=7.8in
\epsffile{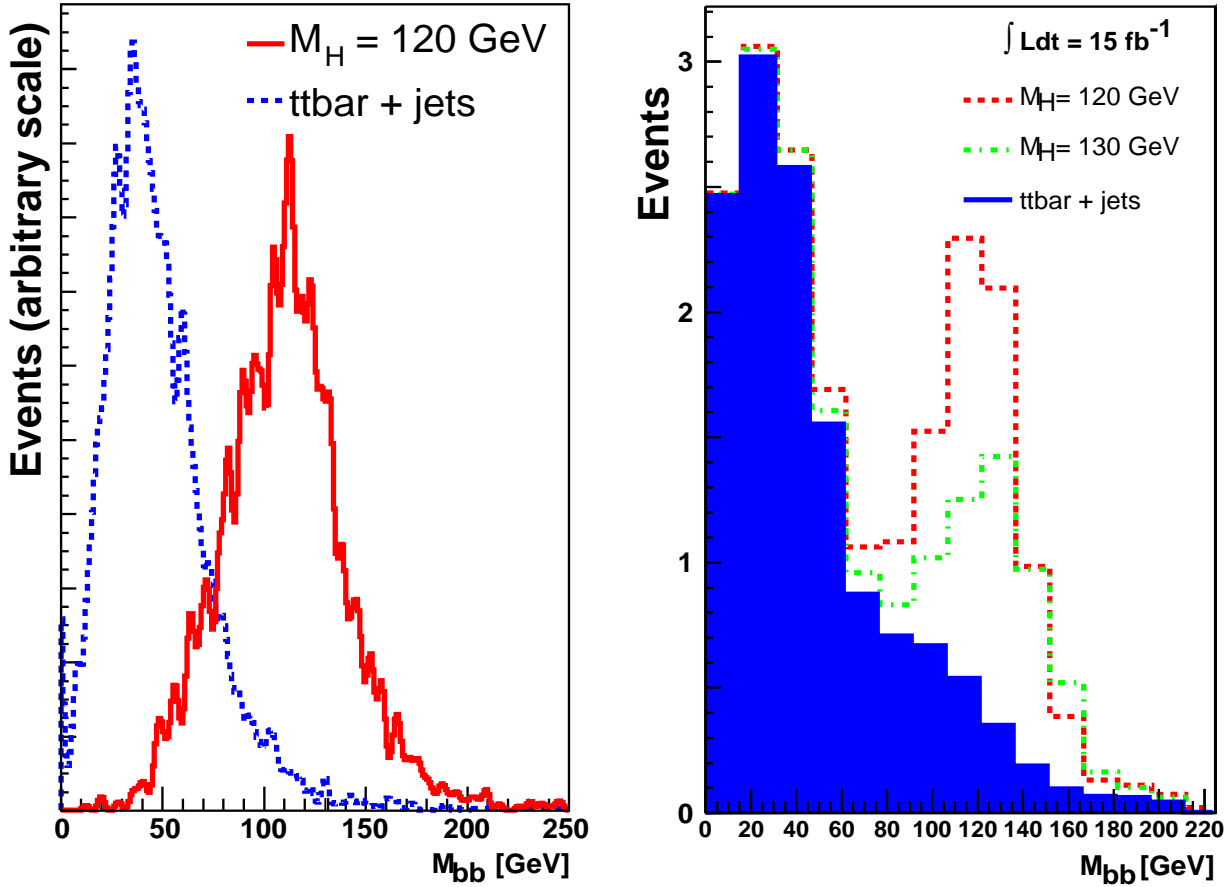}
\end{center}
\vskip -3.5in
\caption{(a) The $t\bar{t}H$ signal (solid) and $t\bar{t}+jets$ (including 
$t\bar{t}b\bar{b},t\bar{t}c\bar{c}$) background (dotted) distributions for 
the fourth-highest mass pair of $b$-tag candidates. 
This includes cases where a jet from the hadronic $W$ decay is misidentified 
as a $b$ jet. We cut on this distribution at $m_{bb} > 60$~GeV.
(b) Invariant mass distribution of $b$-tag candidates after top quark pair 
reconstruction with efficiency $\epsilon_{rec} = 70\%$. For $M_H = 120$~GeV 
and 15~fb$^{-1}$ this corresponds to a $2.8\sigma$ significance for one 
experiment, or $4.1\sigma$ for two.
}
\label{fig:mass}
\end{figure*}

To extract these few signal events we study their reconstruction by performing 
a generator level Monte Carlo simulation using Pythia 6.115, which 
is in good agreement with our parton-level calculations for $t\bar{t}b\bar{b}$ 
production using a jet separation corresponding to a parton-level cone cut of 
$\triangle R > 0.7$. 
Event selection is determined using a parameterized detector simulation. 
For reasonable rejection of the $t\bar{t}+jets$ background, 
we require at least one isolated lepton and 4 jets with $p_T > 15$~GeV, 
at least 3 of which are $b$ tagged, and two additional jets with $p_T > 10$~GeV
as well as a missing $E_T > 15$~GeV. 
The $t\bar{t}b\bar{b}$ ($t\bar{t}c\bar{c}$) 
background differs from signal primarily in that the 
invariant mass of the $b$ ($c$) 
quarks from gluon splitting will be quite low, whereas 
the $b$ quarks from the Higgs boson will have a much higher invariant mass. 
Prior to top reconstruction, however, 
in any given signal (background) event we 
do not know which of the four $b$ jet 
candidates are from the Higgs boson (gluon). 
Thus, for each event we form invariant masses of all combinations of these $b$ 
jet candidates and order them. 
We find there is considerable separation between 
signal and background in the fourth-highest of these ordered dijet masses, and 
cut at 60 GeV in this distribution, see Fig.~\ref{fig:mass}(a). 
Incorrect 
selection of a jet from the hadronic $W$ decay is taken into account.

In Run I, $t\bar{t}$ reconstruction efficiencies of $60\%$ 
were achieved by CDF 
for double $b$-tagged events~\cite{effic}. 
Studies indicate that this can be enhanced by improvements in jet-parton 
assignments resulting from better tracking and energy corrections expected 
in Run II.
Further improvements may even be possible via detector enhancements to $b$ 
flavor tagging. 
However, the efficiency for $t\bar{t}H$ may be lower than for $t\bar{t}$, 
but we do not make the distinction here.
In this paper we will use an optimistic $70\%$ efficiency 
but note that results are only 
slightly worse for $60\%$. 
Although this is somewhat higher than that of the LHC studies, \cite{LHC_TDR},
the jet multiplicity at the Tevatron is significantly lower than at the LHC, 
and as the $t\bar{t}H$ events would be produced closer to threshold at the 
Tevatron, they would be better measured than at the LHC.
Using $70\%$ efficiency we plot in Fig.~\ref{fig:mass}(b) the invariant mass 
distribution of the additional jets after top quark pair reconstruction. 
Here the signal exhibits a significant peak above the background.
For $M_H = 120$~GeV, this would correspond to a $2.8\sigma$ observation for 
one experiment using Poisson statistics converted to a Gaussian equivalent, 
or $4.1\sigma$ for two experiments. Increasing the integrated luminosity to 
20~fb$^{-1}$ increases the significances to $3.3\sigma / 4.7\sigma$, 
respectively.

For higher Higgs boson masses, $M_H \geq 140$~GeV, the dominant signature is 
$W^+W^-W^+W^- b\bar{b}$.
One possible strategy is to demand $\geq 3$ charged leptons, which is one of 
the smaller components of this decay mode at $\sim 6\%$ of the total rate. 
For $M_H = 160$~GeV and 15~fb$^{-1}$ of data for one experiment, this 
translates to only 3.0 events of the 45 total $t\bar{t}H$ events produced. 
Adding a requirement to see $\geq 1$ $b$ tag and minimal cuts, the signal 
sample is only 2 events. The expected background, however, is much less than 
one event, so there is the potential to observe one or two spectacular events 
per experiment\footnote{Any background from $t\bar{t}W$ may be eliminated by 
requiring the observation of extra hadronic activity from the fourth signal 
$W$-boson.}. 
This estimate holds approximately over the Higgs boson mass range 140-170~GeV. 

The signal sample containing exactly two charged leptons is larger, 
by a factor of 3.6, but there is much more background. 
Requiring the two leptons to have the same charge reduces the background 
dramatically, to of order 2 events if additional jets 
are required to be observed.
Further mass constraints may make this much smaller than one.
However, since only 1/3 of the two lepton signal sample will have same-sign 
leptons, one expects another 3-4 spectacular events per experiment. 

Despite the low signal rate for $t\bar{t}H$ associated production events, there 
is considerable potential in this channel to observe a SM Higgs boson, or to set 
exclusion limits on its mass. We have not examined cases of non-SM Higgs bosons, 
but the outlook is optimistic as production cross sections times branching ratio 
are often enhanced relative to the SM, e.g. in SUSY or topcolor models. 
Our analysis suggests an optimistic search strategy for Higgs bosons up to 
about 140~GeV in the $H\to b\bar{b}$ channel, and some promise in the 
$H\to W^+W^-$ channel for the mass range 140-170~GeV.
A thorough understanding of and improvements to the top quark 
reconstruction efficiency and the identification of $b$, $c$ and light quark
jets are very important for this search.  
Increasing the integrated luminosity of the Tevatron by a factor of three would 
increase the significance of this search to $5\sigma$ per experiment.

\acknowledgements
We would like to thank 
U.~Baur, E.~Eichten, K.~Ellis, C.~T.~Hill, M.~Mangano and C.~Quigg 
for useful discussions. 
Fermilab is operated by URA under DOE contract No.~DE-AC02-76CH03000.

\tighten

\end{document}